\documentstyle{article}
\title{
\hspace*{5.3cm} {\large gr-qc/9503056, IMAFF-RC-03-95\vspace*{2.9cm}}\\
\Large{{\bf Canonical Quantization of the Belinski\v{\i}-Zakharov
One-Soliton Solutions\vspace*{1.8cm}}}}
\author{\\Nenad Manojlovi\'c\vspace*{.8cm}
\\Universidade do Algarve,
Unidade de Ciencias Exactas e Humanas,\\
Campus de Gambelas, 8000 Faro, Portugal,
\vspace*{1.3cm}\\ and\vspace*{1.3cm}
\\Guillermo A. Mena Marug\'an\vspace*{.8cm}\\
Instituto de Matem\'aticas y F\'{\i}sica
Fundamental, C.S.I.C.,\\ Serrano 121, 28006 Madrid, Spain.
\vspace*{2.cm}\\ }
\date{\empty}

\begin{document}

\maketitle
\large
\setlength{\baselineskip}{.825cm}

\newpage

\vspace*{1.5cm}

\begin{center}
{\bf Abstract}
\end{center}

\vspace*{.24cm}

We apply the algebraic quantization programme proposed by Ashtekar
to the analysis of the Belinski\v{\i}-Zakharov classical spacetimes,
obtained from the Kasner met\-rics by means of a generalized soliton
transformation. When the solitonic parameters associated with this
transformation are frozen, the resulting Belinski\v{\i}-Zakharov metrics
provide the set of classical solutions to a gravitational
minisu\-per\-space model
whose Einstein equations reduce to the dynamical equations gener\-ated by
a homogeneous Hamiltonian constraint and to a couple of second-class
con\-straints. The reduced phase space of such
a model has the symplectic structure of the cotangent bundle
over $I\!\!\!\,R^+\times I\!\!\!\,R^+$. In this reduced phase space,
we find a complete set of real observables which form a Lie algebra under
Poisson brackets. The
quantization of the gravitational model is then carried out by constructing
an irreducible unitary representation of that algebra of observables.
Finally, we show that the quantum theory obtained in this way is unitarily
equivalent to that which describes the quantum dynamics of the
Kasner model.
\vspace*{1.7cm}


\newpage
\renewcommand{\thesection}{\arabic{section}.}
\renewcommand{\theequation}{\arabic{section}.\arabic{equation}}
\renewcommand{\thefootnote}{a}

\section {Introduction}

In recent years, a special attention has been paid in General Relativity
to the search for exact solutions possessing two commuting Killing
vectors.$^{1,2}$ Spacetimes of this kind provide exact gravitational
solutions in a variety of interesting physical problems, eg, in models with
plane, cylindrical or stationary axial symmetry.$^1$

The Einstein field equations for spacetimes with a two-dimensional
Abelian group of isometries which act orthogonally and transitively on
non-null orbits are non-linear partial differential equations in two
variables.$^3$ A nice property of such field equations is that they admit
symmetry transformations which leave invariant the set of all
classical solutions. Using this result, a series of solution-generating
tech\-niques have been developed to construct new exact
solutions from known ones in models with an Abelian two-parameter
group of isometries.$^2$ Among these tech\-niques, prob\-ably the most
powerful generating method is the inverse-scattering
transformation of Belinski\v{\i} and Zakharov,$^4$ both
because of its simplicity and generality.

The Belinski\v{\i}-Zakharov (BZ) technique, or soliton transformation, is a
general\-ization of the inverse-scattering method which has proved to be so
fruitful in ana\-lysing non-linear partial differential equations in two
dimensions exhibiting so\-liton solutions.$^5$ This generalization
consists essentially in substituting the fixed poles of the
inverse-scattering transformation by pole trajectories. In particular,
Belinski\v{\i} and Zakharov applied this generalized soliton transformation
to the Kasner metric$^3$ (ie, to the diagonal Bianchi type I metric)
to obtain a new exact solution of the Einstein equations in vacuum which
is now known as the BZ one-soliton solution,$^4$ and which can be described
by the non-diagonal inhomogeneous metric
\begin{equation} ds^2=f(t,z)\,(-dt^2+dz^2)\,+\,g_{ab}(t,z)\,dy^a dy^b,
\,\,\,\,\,\,\,\,(a,b=1,2),\end{equation}
\begin{equation} f(t,z)=C\,t^{2p^2-1}\,\frac{\cosh{(pr+D)}}
{|\sinh{(\frac{r}{2})}|},\end{equation}
\begin{equation}
g_{11}(t,z)=AB\,t^{1+2p}\,\frac{\cosh{\left(\{\frac{1}{2}+p
\}r+D\right)}}{\cosh{(pr+D)}},\end{equation}
\begin{equation}
g_{22}(t,z)=\frac{A}{B}\,t^{1-2p}\,\frac{\cosh{\left(\{\frac{1}{2}-p
\}r-D\right)}}{\cosh{(pr+D)}},\end{equation}
\begin{equation}
g_{12}(t,z)=-A\,t\,\frac{\sinh{(\frac{r}{2})}}{\cosh{(pr+D)}}.
\end{equation}
Here, $t$, $z$ and $y^a$ ($a=1,2$) are a set of spacetime coordinates,
$A$, $B$ and $C$ are positive constants, $p$ and $D$ are
real parameters, and $r$ is defined as
\begin{equation} r=\ln{\left[\left(\frac{\mu}{t}\right)^2\right]},\,\,\,\,
\,\,\,\,\,\mu=(z_0-z)-\sqrt{(z_0-z)^2-t^2},\end{equation}
$\mu$ being the pole trajectory of the BZ transformation$^1$, and
$z_0$ a real constant that can be interpreted as the origin of the
$z$ coordinate.

The BZ metric (1.1-6) is defined, in principle, only for $(z-z_0)^2
\geq t^2$, so that $\mu$ in (1.6) be real. Nevertheless, the above
solution can be extended to the region $(z-z_0)^2<t^2$ for some particular
values of the parameters appearing in expressions (1.2-5), like, for
instance, for $p=\frac{1}{2}$.$^6$

Our aim in this work is the construction of a consistent quantum
description for the BZ one-soliton model. The motivation is twofold.
On the one hand, the analysis to be presented can be understood as a
previous step before dealing with the quantization of inhomogeneous
gravitational models admitting two commuting spacelike Killing fields
and whose physical degrees of freedom depend not only on time, but also
on one spatial coordinate. On the other hand, we want to discuss whether
the relation between different families of classical spacetimes which is
provided
by the soliton transformation has any quantum mechanical counterpart.
The quantization of the BZ model will supply an explicit example
for the study of this issue, because the Kasner metric (the seed metric that
leads to the BZ solutions through the soliton transformation$^{1,4}$)
has already been quantized in the literature successfully.$^{7,8}$

To quantize the BZ one-soliton solutions, we
will apply the extension of Dirac's canonical quantization programme$^9$
proposed by Ashtekar {\it et al}.$^{10,11}$
We will first identify
the physical degrees of freedom of the considered model and determine the
symplectic structure of the associated reduced phase space. Using this
structure, we will find a complete set of real classical observables
that form a Lie algebra under Poisson brackets. We will then repres\-ent
these observables by quantum operators acting on a particular vector
space, each element of this space representing a physical quantum state.
Finally, we will fix the inner product in the space of physical states
by imposing a set of reality conditions,$^{11-13}$ that is, by promoting to
adjointness requirements on quantum operators the complex conjugation
relations that exist between the classical observables of the system.

In this way, apart from being of interest by the reasons explained
above, the quantization of the BZ solutions will provide
a new example to be added to the now relatively large list of minisuperspace
gravitational models in which it has been possible to check
the consistency and applicability of the non-perturbative canonical
quantization programme ellaborated by Ashtekar.$^{7,8,14}$

The remainder of this paper is organized as follows. We briefly discuss
the general form of the Einstein field equations for spacetimes with
two commuting spacelike Killing vectors in Sec. 2, where we also review
some results on the soliton transformation. In Sec. 3 we re-examine
the quantization of the diagonal Bianchi type I, following as
close as possible the quantization methods that are to be em\-ployed
in the study of the BZ one-soliton metrics. In Sec. 4
we prove that the Einstein equations for the family of BZ  one-soliton
solutions
can be interpreted as the dynamical equations generated by a homogeneous
gravitational Hamilton\-ian, supplemented with a set of second-class
constraints.
The corresponding global structure of the reduced phase space of the BZ
model is determined in Sec. 5. In that section, we also carry out to
completion the quantization of the studied grav\-itational system, and
compare the obtained quantum theory with that constructed in Sec. 3
for the Kasner model. Finally, we present our conclusions in Sec. 6.

\section{The BZ One-Soliton Transformation}
\setcounter{equation}{0}

For spacetimes possessing two commuting spacelike Killing vectors,
the four-dimen\-sio\-nal metric can always be expressed in the generic
form (1.1). Defining then
\begin{equation} |g|={\rm det}\left\{g_{ab}(t,z)\right\},\end{equation}
one can show that the Einstein equations imply that $|g|^{\frac{1}{2}}$
must satisfy the following wave equation in two dimensions$^1$
\begin{equation} \partial_t^2|g|^{\frac{1}{2}}-
\partial_z^2|g|^{\frac{1}{2}}=0.\end{equation}
As a consequence, the time coordinate can
always be chosen proportional to $|g|^{\frac{1}{2}}$,
\begin{equation} |g|^{\frac{1}{2}}\propto t.\end{equation}
As long as $|g|$ depends only on time, the
$(t,t)$ and $(t,z)$ components of the Einstein equations for the metrics
(1.1) can be seen to adopt the respective expressions
\begin{equation}
G_t^t=\frac{1}{4f}\left(\partial_t\ln{f}\,\partial_t\ln{|g|}+
\frac{\partial_t^2|g|}{2|g|}-\frac{1}{2}g^{ab}(\partial_t^2g_{ab}+
\partial_z^2g_{ab})\right)=0,\end{equation}
\begin{equation} G_t^z=-\frac{1}{4f}\left(\partial_z\ln{f}\,\partial_t
\ln{|g|}-g^{ab}\partial_z\partial_tg_{ab}\right)=0,\end{equation}
where $g^{ab}$ is the inverse of the metric $g_{ab}$ and the lower
case Lattin letters from the beginning of the alphabet denote spatial
indices, with values equal to 1 or 2.

The rest of non-vanishing components of the Einstein equations turn out to
be equivalent to the integrability conditions for the system (2.4,5),
and can be written in the compact form$^1$
\begin{equation} \partial_t\left(A_a^b\right)-\partial_{z}\left(B_a^b
\right)=0,\end{equation}
with
\begin{equation} A_a^b=t\,\partial_tg_{ac}\,g^{cb}\,\,\,\,\,\,\,{\rm and}
\,\,\,\,\,\,\,B_a^b=t\,\partial_zg_{ac}\,g^{cb}.\end{equation}

The BZ generating technique exploits the fact that the non-linear
system (2.6) can be regarded as well as the integrability conditions
associated with the linear eigenvalue problem$^{1,15}$
\begin{equation} \left(\partial_t-\frac{2\lambda t}{\lambda^2-t^2}
\partial_{\lambda}\right)\Psi_{ab}=-\,\frac{t A_a^{c}+\lambda
B_a^c}{\lambda^2-t^2}\,\,\Psi_{cb},\end{equation}
\begin{equation} \left(\partial_z-\frac{2\lambda^2}{\lambda^2-t^2}
\partial_{\lambda}\right)\Psi_{ab}=-\,\frac{\lambda A_a^{c}+t
B_a^c}{\lambda^2-t^2}\,\,\Psi_{cb},\end{equation}
$\lambda$ being a complex variable and $\Psi(t,z,\lambda)$ a $2\times 2$
matrix. If we know a particular solution $(f^{(0)}(t,z),g^{(0)}(t,z))$
of the system of equations (2.2) and (2.4-6)
(what we call a seed metric), the
resolution of the eigenvalue problem (2.8,9), for $A_a^b$ and $B_a^b$
evaluated at $g^{(0)}$ and with boundary condition
\begin{equation} \lim_{\lambda\rightarrow 0} \Psi_{ab}(t,z,\lambda)=
g^{(0)}_{ab}(t,z)\,,\end{equation}
allows us to obtain a new solution to the Einstein equations in the
following way. We first define the pole trajectory $\mu$ as on the
right hand side of Eq. (1.6). Then, the one-soliton transform of the
seed metric $(f^{(0)}(t,z),g^{(0)}(t,z))$ can be calculated by the
formulae$^{1,15}$
\begin{equation} f=Nf^{(0)}\frac{\mu^2 Q}{\sqrt{|t|}(\mu^2-t^2)},\,\,
\,\,\,\,\,\,\,g_{ab}=\left|\frac{\mu}{t}\right|\left(g^{(0)}_{ab}
-\frac{(\mu^2-t^2)}{\mu^2 Q} L_a L_b\right),\end{equation}
with
\begin{equation} L_a=m_c\left(\left.\Psi^{-1}\right|_{\lambda=\mu}
\right)^{cb} g^{(0)}_{ba},\,\,\,\,\,\,\,\, \,\,\,\,\,\,\,\,
Q=L_a g^{(0)\,ab}L_b.\end{equation}
Here, $(m_1,m_2)$ are two complex parameters, $N$ is a constant
and $(\Psi^{-1}|_{\lambda=\mu})^{cb}$ is the inverse of the
matrix $\Psi_{ab}$ [solution to Eqs. (2.8-10)] evaluated at $\lambda=\mu$.

In particular, one can take the Kasner metrics as the seed for the
above transformation, since,
with an appropriate choice of the time gauge, these metrics
can be written in the form (1.1), with
metric functions $f$ and $g$ given by$^{15}$
\begin{equation} f= \tilde{C}\, t^{2p^2-\frac{1}{2}},\,\,\,\,\,\,\,
\,\,\,\,\,\,\,\,\,\,g_{11}=AB\, t^{1+2p},\end{equation}
\begin{equation} g_{22}=\frac{A}{B}\,t^{1-2p}\,,\,\,\,\,\,\,\,\,\,\,\,
\,\,\,\,\,\,\,\,g_{12}=0\,.\,\,\,\,\,\,\end{equation}
$A$, $B$ and $\tilde{C}$ are three positive constants and
$p\in I\!\!\!\,R$. For real solitonic parameters
$(m_1,m_2)$, the new exact vacuum solutions to the Einstein equations
that one reaches in this way are precisely the BZ
inhomogeneous metrics displayed in Eqs. (1.1-6).

\section{The Diagonal Bianchi Type I Model}
\setcounter{equation}{0}

Although the canonical quantization of the diagonal Bianchi type I model
has already been completed,$^{7,8}$ we want to present here a
slightly different version of the quantization procedure which will prove
specially suited to discuss the relation be\-tween the quantum theories
that respectively describe the BZ one-soliton metrics and the Kasner
solutions. We will begin by analysing the classical dynamics
of the Kasner model in Subsection 3.a, and attain the desired
quantization in Subsection 3.b.
\vspace*{.4cm}

\noindent{\bf 3.a. Classical Analysis}
\vspace*{.3cm}

The Kasner metrics (2.13,14) admit the generic expression
\begin{equation} f=e^{2Z(t)},\,\,\,\,\,\,\,\,\,\,\,\,
\,g_{11}=e^{2X(t)+2Y(t)},\end{equation}
\begin{equation}g_{22}=e^{2X(t)-2Y(t)},\,\,\,\,\,\,\,\,\,\,\,\,\,\,\,
g_{12}=0.\,\,\,\,\,\,\end{equation}
{}From now on, we will regard these equations as the definition of a
homogeneous minisuperspace model whose degrees of freedom are just the
functions $X$, $Y$ and $Z$. The determinant
of the metric $g_{ab}$ depends thus only on time:
$|g|=\exp{\{4X(t)\}}$.

The $(t,z)$ component of the Einstein equations, given by formula (2.5),
is ident\-ically zero for metrics (3.1,2). From
Eq. (2.4), on the other hand, the $(t,t)$ component of the Einstein
equations, which can be interpreted
as the gravitational Hamiltonian constraint at each point of the spacetime,
reduces in this case to
\begin{equation} {\cal H}\equiv G_t^t=\frac{1}{f}\left[2\frac{dX}{dt}
\frac{dZ}{dt}+\left(\frac{dX}{dt}\right)^2-\left(\frac{dY}{dt}\right)
^2\right]=0.\end{equation}
The integration of this Hamiltonian constraint over each surface
of constant time gives the total Hamiltonian, H, that generates
the dynamics of the studied grav\-itational model. To simplify calculations,
it is convenient to choose the following time gauge:
\begin{equation} dt=|g|^{\frac{1}{2}}\,dT=e^{2X(T)}\,dT.\end{equation}
The Hamiltonian $H$ that dictates the evolution in the new time coordinate
$T$ can then be defined as
\begin{equation} H=\frac{1}{V_{\Omega}}\int_{\Omega}\,\sqrt{|g^{(4)}|}\,
{\cal H}=[2\dot{X}\dot{Z}+\dot{X}^2-\dot{Y}^2].\end{equation}
Here, $\int_{\Omega}$ denotes integration over a constant-$T$ section of
the spacetime, $V_{\Omega}$ is the volume of that section ($V_{\Omega}=
\int_{\Omega}1$), $g^{(4)}$ is the determinant of the 4-metric, and
the dots represent the derivative with respect to $T$. The
right hand side of formula (3.5) has been obtained by using Eqs. (3.3,4)
and the fact that, for the set of coordinates $(T,z,y^1,y^2)$,
\begin{equation} \sqrt{|g^{(4)}|}=f\,e^{4X}.\end{equation}
It is worth noticing that the Hamiltonian $H$, as given by Eq. (3.5), is
well-defined in the limit $V_{\Omega}\rightarrow\infty$.

{}From the above Hamiltonian, it is possible to deduce the expressions of
the momenta canonically conjugated to $X$, $Y$ and $Z$ by assuming
the implicit dependence of the first time derivatives $\dot{X}$,
$\dot{Y}$ and $\dot{Z}$ on such momenta. Let us first
introduce the notation $u_i\equiv\{X,Y,Z\}$ $(i=1,2,3)$. Differentiating
then H with respect to $\dot{u}_i$, applying the chain rule and
substituting the Hamiltonian equations $\partial_{p_i}H=\dot{u}_i$,
we get
\begin{equation}\partial_{\dot{u}_i}H=\partial_{\dot{u}_i}p_j\,
\partial_{p_j}H=\partial_{\dot{u}_i}p_j\,\dot{u}_j,\end{equation}
where $p_j$ is the momentum conjugate to $u_j$. Making use of formulae
(3.5) and (3.7), it is now a simple exercise to show that
\begin{equation} p_X=2(\dot{X}+\dot{Z}),\,\,\,\,\,\,\,\,\,\,p_Y=-2\dot{Y},
\,\,\,\,\,\,\,\,\,\,p_Z=2\dot{X},\end{equation}
up to irrelevant additive constant factors. Therefore, the Hamiltonian
(3.5) can be rewritten in terms of the canonical momenta as
\begin{equation} H=\frac{1}{4}[2p_Xp_Z-p_Z^2-p_Y^2].\end{equation}

The dynamical equations that follow from this Hamiltonian state that
$p_X$, $p_Y$ and $p_Z$ are constants in the evolution, and that $\dot{X}$,
$\dot{Y}$ and $\dot{Z}$ are given by Eqs. (3.8). In particular,
recalling that $|g|^{\frac{1}{2}}=\exp{(2X)}$ and Eq. (3.4), we have
\begin{equation} \frac{d|g|^{\frac{1}{2}}}{\,dt\,}=2\dot{X}=p_Z=
{\rm const.}\end{equation}
Thus, $|g|^{\frac{1}{2}}$ turns out to be linear in the time coordinate $t$,
and the Einstein equation (2.2) is straightforwardly satisfied.
Moreover, it is not difficult to check that, for the metrics
(3.1,2), the system of non-linear equations (2.6) is already contained
in the equations of motion implied by the Hamiltonian (3.9). This is
not surprising because, being the $(t,z)$ component
of the Einstein tensor identically vanishing in our case, Eqs. (2.6)
simply provide the integrability conditions for the constraint
(3.3) and, hence, for the Hamiltonian (3.9).
So, the Einstein equations for the minisuperspace model
analysed here are equivalent to the dynamics generated by the homogeneous
Hamiltonian (3.9) and to the first-class constraint $H=0$.

Since $p_X$, $p_Y$ and $p_Z$ are constants, Eqs. (3.8) can be immediately
integrated to give the classical solutions of the model.
By choosing the origin of the time coordinate $T$ in such a way that
\begin{equation} X(T=0)=0,\end{equation}
we get
\begin{equation} X(T)=\frac{p_Z}{2}\,T,\,\,\,\,\,\,\,\,\,
Y(T)=-\frac{p_Y}{2}\,T+Y_0,\,\,\,\,
\,\,\,\,\,Z(T)=\frac{p_X-p_Z}{2}\,T+Z_0,\end{equation}
where $Y_0$ and $Z_0$ are two integration constants, and $(p_X,p_Y,p_Z)$
must satisfy the constraint $H=0$.

If we want to consider only different classical 4-geometries, it is
necessary to restrict the allowed range of the parameters on which
the classical solutions depend. We notice that the
4-geometries obtained from Eqs. (3.1,2), (3.4) and (3.12) are invariant
under time reversal if one also flips the signs of the canonical
momenta $(p_X,p_Y,p_Z)$.$^8$ In order to elliminate this redundancy, we
can restrict to the sector of solutions with
\begin{equation} p_Z=2\dot{X}(T)\in I\!\!\!\,R^+\,\end{equation}
which contains all possible 4-geometries. On the other hand,
if we assume that the coordinates $y^1$ and $y^2$ that appear in Eq. (1.1)
are physically indistinguishable, the metrics $g_{ab}$ related by an
interchange of the indices 1 and 2 describe
the same 4-geometry. From Eqs. (3.1,2) and (3.12), we must then identify
those classical solutions which differ just in the sign of the parameters
$p_Y$ and $Y_0$. To take into account each physical solution only once,
we will restrict $p_Y$ to be a negative constant from now on.
Thus, we will impose the following ranges for the parameters
present in Eq. (3.12):
\begin{equation} Y_0,\,Z_0\in I\!\!\!\,R\,,\,\,\,\,\,\,\,\,\,\,
p_Y\in I\!\!\!\,R^-\,,\,\,\,\,\,
\,\,\,\,\,p_Z\in I\!\!\!\,R^+\,,\end{equation}
whereas $p_X$ is fixed by $p_Y$ and $p_Z$ through the constraint $H=0$:
\begin{equation} p_X=\frac{p_Z}{2}+\,\frac{p_Y^2}{2p_Z}\,.\end{equation}

Note that, being $\dot{X}$ a positive constant, we can explicitly define
the time coordinate $t$ in Eq. (3.4) as
\begin{equation} t=\int^T_{-\infty}\,e^{2X(T)}\,dT,\end{equation}
so that, on the classical solutions (3.12), the relation between the two
time gauges employed in our calculations is given by
\begin{equation} e^{p_ZT}=p_Zt.\end{equation}
If one introduces now the redefinitions
\begin{equation} p_Z=A,\,\,\,\,\,\,\,\,p_Y=-2pA,\,\,\,\,
\,\,\,\,Y_0=\frac{1}{2}\ln{B}-p\ln{A},\end{equation}
\begin{equation} Z_0=\frac{1}{2}\ln{\tilde{C}}+\left(\frac{1}{4}-p^2
\right)\ln{A},\end{equation}
where, from conditions (3.14),
\begin{equation} A,\,B,\,\tilde{C},\,p\in I\!\!\!\,R^+,\end{equation}
a trivial computation shows that the classical metrics determined by
Eqs. (3.1,2), (3.12) and (3.15) are just the Kasner
metrics (2.13,14), with the parameter $p$ restricted to be positive
to take into account the above discussed symmetry under
interchange of coordinates.
\vspace*{.4cm}

\noindent{\bf 3.b. Quantum Analysis}
\vspace*{.3cm}

We can now proceed to the quantization of
the diagonal Bianchi type I model. Our first step will consist in
determining the symplectic structure of the reduced phase space of this
gravitational system. This structure can be obtained as the pull-back
to the constraint surface $H=0$ of the symplectic form of the unreduced
phase space of the model,$^{16}$
\begin{equation} \left.\Gamma=\left. dX\wedge dp_X+dY\wedge dp_Y +dZ
\wedge dp_Z \right. \right|_{H=0},\end{equation}
where we have used the fact that $(X,Y,Z,p_X,p_Y,p_Z)$ form a canonical
set of phase space variables.

The symplectic form (3.21) can be proved to be time-independent. Therefore,
we can evaluate it at any constant-$T$ section
of the spacetime, the final result being insensitive to the specific section
selected. Choosing the $T=0$ surface and recalling condition
(3.11), we get
\begin{equation} \Gamma=dY_0\wedge dp_Y+dZ_0\wedge dp_Z,\end{equation}
with $Y_0$ and $Z_0$ the initial values of $Y$ and $Z$ [see Eq. (3.12)].
The change of variables (3.18,19) then leads to
\begin{equation} \Gamma= dA\wedge dP_A + dp\wedge dP_p,\end{equation}
where
\begin{equation} P_A=p\ln{B}-\frac{1}{2}\ln{\tilde{C}},\,\,\,\,\,\,\,\,
\,\,\,P_p=A\left(\ln{B}-2p\right),\end{equation}
or, equivalently,
\begin{equation} B= e^{2p+\,\frac{P_p}{A}},\,\,\,\,\,\,\,\,\,\,\,\,\,
\tilde{C}=e^{4p^2+2\,\frac{pP_p}{A}\,-2P_A}.\end{equation}
Equations (3.20) and (3.24) imply that
\begin{equation}
A,\,p\in I\!\!\!\,R^+\,\,\,\,\,\,\,\,\,{\rm and}\,\,\,\,\,
\,\,\,\,P_A,\,P_p\in I\!\!\!\, R,\end{equation}
so that we can interpret the symplectic structure obtained for the reduced
phase space of the diagonal Bianchi type I model as that
corresponding to the cotangent bundle over $I\!\!\!\,R^+\times
I\!\!\!\,R^+$.

A complete set of elementary variables in this reduced phase space
(ie, a complete set of observables) is provided by $A$, $p$ and their
generalized momenta$^8$
\begin{equation} L_A=AP_A\,,\,\,\,\,\,\,\,\,\,\,\,\,\,L_p=pP_p.
\end{equation}
Conditions (3.26) ensure that all these observables are real.
On the other hand, since their only non-vanishing Poisson brackets are
\begin{equation} \{A,L_A\}=A\,,\,\,\,\,\,\,\,\,\,\,\,\,\,
\{p,L_p\}=p,\end{equation}
they form the Lie algebra $L(T^{\ast}GL(1,I\!\!\!\,R)\times
T^{\ast}GL(1,I\!\!\!\,R))$, $T^{\ast}GL(1,I\!\!\!\,R)$ being the semidirect
product of $I\!\!\!\,R$ and $I\!\!\!\,R^+$.

In order to quantize the model,
we will represent the observables $(A,p,L_A,L_p)$
as operators acting on the vector space of complex functions $\Psi(A,p)$
over $I\!\!\!\,R^+\times I\!\!\!\,R^+$. Each complex function $\Psi$
will represent in this way a physical quantum state of the diag\-onal
Bianchi type I model, because we have already got rid of all the
constraints of the system. The action of the operators
$(\hat{A},\hat{p},\hat{L}_A,\hat{L}_p)$
can be explicitly defined as
\begin{equation} \hat{A}\Psi(A,p)=A\, \Psi(A,p),\,\,\,\,\,\,\,\,\,\,\,\,\,
\,\hat{p}\Psi(A,p)=p\,\Psi(A,p),\end{equation}
\begin{equation} \hat{L}_A\Psi(A,p)=-i A\partial_A\Psi(A,p),\,\,\,
\,\,\,\,\,\,\hat{L}_p\Psi(A,p)=-i p\partial_p\Psi(A,p),\end{equation}
where we have taken $\hbar=1$.
It is easy to check that the above operators form
a closed algebra under commutators which coincides
with the Lie algebra of Poisson brackets (3.28).

According to Ashtekar,$^{11}$ the inner product in the space of physical
states can be determined by imposing a set of reality conditions.$^{12}$
For the elementary observables that we have chosen, the reality conditions
demand that the operators (3.29,30) be self-adjoint, because they all
represent real classical observables. These hermiticity requirements
select the physical inner product
\begin{equation} <\Phi,\Psi>=\int_{I\!\!\!\,R^+\times I\!\!\!\,R^+}
\,\frac{dA}{A}\frac{dp}{p}\,\Phi^{\ast}(A,p)\Psi(A,p),\end{equation}
with $\Phi^{\ast}$ the complex conjugate to $\Phi$. So, the Hilbert
space of physical states is
$L^2(I\!\!\!\,R^+\times I\!\!\!\,R^+,dAdp(Ap)^{-1})$.
Actually, what we have obtained by implementing Ashtekar's programme
is just an irreducible unitary representation of the algebra
of observables $L(T^{\ast}GL(1,I\!\!\!\,R)\times T^{\ast}GL(1,I\!\!\!\,R))$.
We notice in this sense that the measure that appears in Eq. (3.31) is
precisely that corresponding
to the product of two copies of the group $T^{\ast}GL(1,I\!\!\!\,R)$.$^{17}$
The results presented in this section are in complete agreement with those
reached in Ref. 7 for the quantization of the Bianchi type I once the
diagonal reduction to the Kasner model is taken into account.

\section{ The BZ One-Soliton Model}
\setcounter{equation}{0}

We turn now to the analysis of the BZ one-soliton solutions, described in
Eqs. (1.1-6). In the following, we will restrict our attention exclusively
to families of BZ metrics with fixed values of the parameters $z_0$ and $D$.
This restriction can be proved to guarantee that the BZ solutions possess
exactly the same number of degrees of freedom as the seed metrics from
which they can be obtained by means of a soliton transformation,$^{15}$
that is, as the diagonal Bianchi type I model. In fact, one can show that
fixing the
constants $z_0$ and $D$ results in freezing all the solitonic parameters
involved in the BZ transformation.$^{15}$

{}From Eq. (1.6), it is obvious that the value of $z_0$ can always be absorbed
by shifting the origin of the $z$ coordinate:
\begin{equation} \tilde{z}=z-z_0.\end{equation}
Restricting $z_0$ to be a given constant is thus equivalent to consider
a unique pole trajectory, determined by
\begin{equation} \mu=-\tilde{z}-\sqrt{\tilde{z}^2-t^2}.\end{equation}
Using this equation and the definiton $r=\ln{[(\mu/t)^2]}$,
one can show that the variable $r$ changes its sign under
the transformation $\tilde{z}\rightarrow -\tilde{z}$. On the other hand,
by ap\-ply\-ing this transformation to the metrics (1.1-5) and flipping the
sign of the spatial coordinate $y^2$, it is straightforward to conclude
that the BZ solutions that differ only in the sign of the constant $D$
turn out to describe the same classical four-dimensional spacetime
geometry. With the aim at keeping this symmetry in our model while
fixing the value of $D$, we will restrict from now on this parameter
to vanish:
\begin{equation} D=0.\end{equation}

The family of BZ solutions to be studied can then
be represented in the form
\begin{equation} f=\frac{\cosh{(\beta(t)r)}}{|\sinh{(\frac{r}{2})}|}
\,e^{2Z(t)},\end{equation}
\begin{equation}g_{11}=\frac{\cosh{\left(\{\frac{1}{2}+\beta(t)
\}r\right)}}{\cosh{(\beta(t)r)}}\,e^{2X(t)+2Y(t)},\end{equation}
\begin{equation}
g_{22}=\frac{\cosh{\left(\{\frac{1}{2}-\beta(t)
\}r\right)}}{\cosh{(\beta(t)r)}}\,e^{2X(t)-2Y(t)},\end{equation}
\begin{equation}
g_{12}=-\,\frac{\sinh{(\frac{r}{2})}}{\cosh{(\beta(t)r)}}\,e^{2X(t)}.
\end{equation}
These equations can be regarded as the definition of a gravitational
minisuperspace model whose degrees of freedom are the functions $X$,
$Y$, $Z$ and $\beta$, which depend only on time.

A simple computation leads to the result
\begin{equation} |g|^{\frac{1}{2}}=e^{2X(t)}.\end{equation}
Choosing now the time coordinate $t$ as in Eq. (2.3), the Einstein equation
(2.2) reduces in our case to the condition
\begin{equation} \frac{dX}{dt}=\frac{1}{2t}\,.\end{equation}
The Einstein equations (2.4,5), on the other hand, provide the two
gravitational constraints that exist in our minisuperspace model. They
can be interpreted in turn as the Hamiltonian constraint and the only
non-vanishing momentum con\-straint at each point of the spacetime.
Realizing that all the $z$-dependence of the metrics (4.4-7)
is contained in the variable $r$, it is easy to check that the momentum
con\-straint $G_t^z=0$ and the linear combination of constraints
\begin{equation} \tilde{G}_t^t\equiv G_t^t+\,
\frac{\partial_t r}{\partial_z r}\,G_t^z=0\end{equation}
determine, respectively, the first partial derivatives
$\partial_r\ln{\tilde{f}}$
and $\partial_t\ln{\tilde{f}}$ of the function
$\tilde{f}(t,r)\equiv f(t,z(t,r))$ once the metric $g_{ab}$
is known. The non-linear equations (2.6), as we have already explained,
are just the integrability conditions for the two con\-straints
of the system. These conditions can be equivalently
imposed by demanding the coincidence
at all points of the second partial derivatives
\begin{equation} \partial_r\partial_t\ln{\tilde{f}}=
\partial_t\partial_r\ln{\tilde{f}}\,,\end{equation}
which can be computed from the constraints $G_t^z=0$ and $\tilde{G}_t^t=0$.
Employing Eq. (4.9), a detailed and lengthy but trivial calculation
shows that the requirement (4.11), and thus Eqs. (2.6), are satisfied if and
only if the following dynamical equations hold
\begin{equation} \left(\frac{dY}{dt}\right)^2
=\left(\frac{\beta}{t}\right)^2,\end{equation}
\begin{equation} \frac{d\beta}{dt}=0.\end{equation}
Finally, substituting Eqs. (4.4-7), (4.9) and (4.12,13) in expression (2.5),
one can prove that there exist no classical solutions with
$\beta=-t(dY/dt)$, whereas for
\begin{equation} \frac{dY}{dt}=\frac{\beta}{t}\end{equation}
the momentum constraint $G_t^z=0$ is trivially satisfied.
We thus conclude that, for the considered minisuperspace model, the whole
set of Einstein equations reduces to the constraint (4.10) and to the
dynamical equations (4.9) and (4.13,14).

Note that Eqs. (4.13,14) determine
the metric function $\beta$ in terms of $Y(t)$.
Our task in the rest of this section will be to demonstrate that these
two equations can be interpreted in fact as second-class constraints which
allow us to elliminate $\beta$ as a physical degree of freedom.

Let us first adopt the same time gauge (3.4) that was used in the analysis
of the diagonal Bianchi type I model. In this gauge,
Eqs. (4.9) and (4.13,14) translate into
\begin{equation} \chi_1\equiv -S(T) \dot{Y}+\beta=0,\end{equation}
\begin{equation} \chi_2\equiv \dot{\beta}=0,\end{equation}
\begin{equation} S(T)\dot{X}=\frac{1}{2},\end{equation}
where the dots denote differentiation with respect to $T$ and
\begin{equation} S(T)=\frac{t(T)}{e^{2X(T)}}\,.\end{equation}

Defining now
\begin{equation} {\cal H}\equiv \tilde{G}_t^t-\frac{1}{f e^{4X}}
\left[\frac{\sinh^2{(\frac{r}{2})}}{S^2(T)\cosh^2{(\beta r)}}\,(S(T)\dot{Y}
+\beta)\,\chi_1+\tanh{(\beta r)}\, r\, \dot{X}\, \chi_2\right],
\end{equation}
a careful calculation, employing Eqs. (4.4-7), (4.10) and (2.4,5), leads
to the result
\begin{equation} {\cal H}=\frac{1}{f\,e^{4X}}
\left[2\dot{X}\dot{Z}+\dot{X}^2-\dot{Y}^2+\frac{1}{4S^2(T)}
-\frac{\sinh^2{(\frac{r}{2})}}{4\cosh^2{(\beta r)}}\,r^2
\dot{\beta}^2\right]\,.\end{equation}

Assuming that $\chi_1$ and $\chi_2$ form a set of second-class constraints,
${\cal H}$, as given by Eqs. (4.10) and (4.19), turns out to be a linear
combination of all the constraints of the system. In that case, the
integration of ${\cal H}$ over each surface of constant time $T$ will
supply us with a total Hamiltonian for the model.$^9$
We will also admit at this point that
Eq. (4.17) is one of the dynamical equations implied by the Hamiltonian
evolution. This assumption allows us to substitute Eq. (4.17) in
formulae (4.15) and (4.20) to get equivalent expressions on shell for
the classical Hamiltonian and the constraint $\chi_1$. In particular,
$\chi_1$ will now read
\begin{equation} \chi_1= \beta-\frac{\dot{Y}}{2\dot{X}}\,\,.\end{equation}
The consistency of the hypotheses introduced in this paragraph will
be proved later on in this section.

Using then Eqs. (4.17) and (4.20), and following a procedure
similar to that explained in Sec. 3 for the Kasner model, we obtain
a total Hamiltonian for the BZ one-soliton model of the form
\begin{equation} H=\frac{1}{V_{\Omega}}\int_{\Omega}\sqrt{|g^{(4)}|}\,
{\cal H}=2\dot{X}\dot{Z}+2\dot{X}^2-\dot{Y}^2-\frac{1}{4} W(\beta)
\dot{\beta}^2,\end{equation}
where
\begin{equation} W(\beta)=\frac{1}{V_{\Omega}}\int_{\Omega}\frac{
\sinh^2{(\frac{r}{2})}}{\cosh^2{(\beta r)}}\,r^2,\end{equation}
and we are employing the same notation as in Sec. 3.a.

Generalizing now the analysis carried out for the diagonal Bianchi type I
case, it is straightforward to deduce the expressions of the momenta
canonically conjugate to $X$, $Y$, $Z$ and $\beta$
from the Hamiltonian (4.22):
\begin{equation} p_X=4\dot{X}+2\dot{Z},\,\,\,\,\,\,\,\,\,p_Y=-2\dot{Y},
\,\,\,\,\,\,\,\,\,p_Z=2\dot{X},\end{equation}
\begin{equation} p_{\beta}=-\frac{1}{2}W(\beta)\dot{\beta}.\end{equation}
Equations (4.16,17) and (4.21,22) can thus be rewritten
\begin{equation} \chi_1= \beta+\frac{p_Y}{2p_Z}=0,\,\,\,\,\,\,\,\,\,\,\,\,
\chi_2= p_{\beta}=0,\end{equation}
\begin{equation} H=\frac{1}{2}p_Xp_Z-\frac{1}{2}p_Z^2-\frac{1}{4}p_Y^2
-\frac{p_{\beta}^2}{W(\beta)},\end{equation}
\begin{equation} p_Z=\frac{1}{S(T)}\,.\end{equation}
The Hamiltonian (4.27), on the other hand, must satisfy the constraint
$H=0$.

It is a simple exercise to check that $\{\chi_1,\chi_2\}=1$ and
that the Poisson brackets of $\chi_1$ and $\chi_2$ with $H$ vanish
weakly (ie, as long as $\chi_2=0$). Therefore, $\chi_1$ and $\chi_2$
form a set of second-class constraints for the system under consideration,
as we wanted to prove. The imposition of these constraints elliminates
the degrees of freedom $(\beta,p_{\beta})$ and leads to a reduced model
with associated Hamiltonian
\begin{equation} H_R=\frac{1}{2}p_Xp_Z-\frac{1}{2}p_Z^2-\frac{1}{4}p_Y^2.
\end{equation}
The Poisson and Dirac brackets of the variables that describe this reduced
model are easily seen to coincide,
for $X$, $Y$, $Z$, $p_X$, $p_Y$ and $p_Z$ commute strongly
with the constraint $\chi_2$.

It is worth remarking that the reduced Hamiltonian (4.29) is
independ\-ent of the topology of the constant-$T$ surfaces. As a
consequence, this Hamiltonian is well-defined in the limit of
non-compact surfaces.

The Hamiltonian equations reached from $H_R$ reproduce the first order
equations (4.24) and assure that $p_X$, $p_Y$ and $p_Z$
are constants of motion. Select\-ing the origin of time by condition
(3.11), the classical solutions of our model turn out then to be
formally identical to those obtained
in Eq. (3.12) for the Kasner metrics, except for that now
\begin{equation} Z(T)=\frac{p_X-2p_Z}{2}\, T +Z_0.\end{equation}
Besides, from the first constraint in Eq. (4.26) we get
\begin{equation} \beta=-\frac{p_Y}{2p_Z}= {\rm const.}\,,\end{equation}
so that $\chi_2=p_{\beta}\propto \dot{\beta}=0$ is automatically satisfied.

If we want to consider only different classical 4-geometries, a
discussion similar to that presented for the diagonal Bianchi type I in
Sec. 3.a leads us to conclude that, owing to the existing symmetry under
time reversal, we must fix the sign of one of the canonical momenta of
our reduced model.
Hence, we will restrict, eg, $p_Z$ to be a positive constant.
On the other hand, it is not difficult to check that an interchange
of the spatial coordinates $y^1$ and $y^2$ in Eq. (1.1) can be
reinterpreted
in the BZ model as a flip of sign in the parameters $p_Y$ and $Y_0$ of
the classical solutions. Therefore, to study all possible 4-geometries,
it will suffice to analyse the sector $p_Y<0$. In this way, the parameters
$(Y_0,Z_0,p_Y,p_Z)$ on which the BZ solutions depend will take on the
range of values displayed in Eq. (3.14). The constant
momentum $p_X$ is determined through the Hamiltonian constraint
$H_R=0$, which implies that
\begin{equation} p_X=p_Z+\frac{p_Y^2}{2p_Z}\,.\end{equation}

Since the expression of the classical solutions $X(T)$ is the same for
the Kasner and the BZ one-soliton metrics, and $p_Z\in I\!\!\!\,R^+$
in both cases, the relation (3.17) between the time coordinates $t$ and $T$
is still valid in the BZ model, provided that we define the coordinate $t$
as in Eq. (3.16).
We then have that, on the classical solutions,
\begin{equation} e^{2X(T)}=e^{p_ZT}=p_Zt.\end{equation}
Substituting this equality in formula (4.18) and making use of the last
of the Hamiltonian equations in (4.24), we arrive at the result
\begin{equation} \frac{1}{S(T)}=p_Z=2\dot{X},\end{equation}
which is precisely Eq. (4.17). So, this equation is simply a consequence
of the dynamical evolution generated by the Hamiltonian (4.29), as we had
anticipated. We thus conclude that the work hypotheses introduced above
Eq. (4.21) are completely consistent.

Finally, adopting the redefinitions (3.18) and
\begin{equation} Z_0=\frac{1}{2}\ln{C}+\left(\frac{1}{2}-p^2\right)
\ln{A},\end{equation}
it is not difficult to prove that the classical metrics obtained from
Eqs. (4.4-7), (4.30-32) and the two first relations in Eq. (3.12)
are nothing else but the family of BZ one-soliton solutions (1.2-5)
with $D=0$ and $p$ restricted to be positive
because of the symmetry under the interchange of
spatial coordinates $y^1$ and $y^2$.

\section{Canonical Quantization of the BZ model}
\setcounter{equation}{0}

We have shown that, in the BZ model, the degree of freedom $\beta(T)$
is determined by a set of second-class constraints. Once these
constraints are elliminated, the rest of degrees of freedom of the system,
described by the canonical set of vari\-ables $(X,Y,Z,p_X,p_Y,p_Z)$,
are still subjected to the Hamiltonian constraint (4.29).
Therefore, to obtain the symplectic structure of the reduced phase space,
we must simply pull-back to the surface $H_R=0$ the symplectic form
\begin{equation} dX\wedge dp_X+dY\wedge dp_Y+dZ\wedge dp_Z\,.
\end{equation}
Given that the desired pull-back is time-independent, we can evaluate it at
any constant-$T$ section of the spacetime. Selecting the $T=0$ surface
and recalling condition (3.11), we arrive at a symplectic form on the
reduced phase space which coincides formally with
that reached in Eq. (3.22) for the diagonal
Bianchi type I. Introducing then the change of variables defined
by Eqs. (3.18) and (4.35),
a simple calculation shows that both the symplectic
form (3.23) and relations (3.24), which were deduced for the Kasner metrics
in Sec. 3.b, are valid in the BZ model with the replacement
of $C$ for $\tilde{C}$. Moreover, since Eqs. (3.14), and hence restrictions
(3.20),\footnote{Substituting $C$ for $\tilde{C}$.} still apply,
the ranges of the canonical variables
$(A,p,P_A,P_p)$ that appear in Eq. (3.23) turn out to be given
again by expression (3.26), as in the diagonal Bianchi type I case.
The symplectic structure of the reduced phase space of the BZ one-soliton
model can thus be identified with that corresponding to the Kasner model,
namely, the symplectic structure of the cotangent bundle over $I\!\!\!\,R^+
\times I\!\!\!\,R^+$.

The quantization of this reduced phase space was already carried out in
Sec. 3.b. All the results presented there [below Eq. (3.26)] hold as well
in the BZ minisuper\-space model studied here.

In this way, we see in particular that the BZ one-soliton
solutions with constant values of $z_0$ and $D$
have the same degrees of freedom as the Kasner metrics. This
is due to the fact that, for fixed solitonic parameters, the BZ
transformation that relates the two mentioned types of classical
solutions preserves the physical degrees of freedom. We recall, in this
sense, that fixing the constants $z_0$ and $D$ is equivalent to
freezing the solitonic parameters of the BZ transformation.

We have also proved that the quantum theories constructed
for the two ana\-lysed models are totally equivalent. The observables
and physical Hilbert spaces of these two quantum theories can
mutually be identified. Note, nevertheless,
that the physical interpretation of these observables in terms of the
metric differs in the two considered models,
because the 4-geometries studied are not the same in each case.
Whether or not the symmetry that exists
between different classical spacetimes related by the BZ transformation
with fixed solitonic parameters translates always into the equivalence
of the quantum theories which describe those gravitational systems is
an issue which deserves further research. The analysis presented
in this work can be understood in this line as the discussion of a
particular example which supports the validity of this conjecture.

We want to close this section with some remarks about the kind of
predictions that can be obtained from the quantum theory that we
have built up for the BZ one-soliton metrics. The gravitational
model employed to describe the BZ one-soliton
solutions possesses as its only degrees of freedom the functions $\beta$,
$X$, $Y$ and $Z$, which depend exclusively on time. Therefore, the
minisuperspace model considered is in fact homogeneous. This explains
why the Hamiltonian (4.29) and the second-class constraints (4.26)
are independent of the spatial coordinates. As a consequence,
all reachable predictions in the corresponding quantum theory refer only
to homogeneous variables, like, for instance, the parameters
$A$, $B$, $C$ and $p$ on which the 3-geometry depends on each
constant-time surface. The quantum theory constructed does not supply us
with an appropriate framework to address questions about local quantities,
such as the expectation value of the Riemann tensor at each point of the
spacetime, a value which would allow us to ellucidate whether the
singularities
of the BZ geometries (1.1-6) disappear or not in the quantum evolution.
To analyse this kind of local problems quantum mechanically, we should
first enlarge our gravitational model to permit the dependence of the
physical degrees of freedom on spatial position. The quantization
of such an enlarged gravitational system would then lead us to a true
quantum field theory.

\section{Conclusions}

Following Ashtekar's programme, we have carried out to completion the
quantization of the family of BZ one-soliton metrics (1.1-6) with
vanishing parameter $D$ and a given constant value of $z_0$. This family
of classical spacetimes can be obtained from the Kasner metrics by means of
a generalized soliton transformation in which the pole trajectory is fixed
and all the solitonic degrees of freedom are frozen.

In order to quantize the BZ one-soliton metrics, we have shown that they
can be regarded as the classical solutions to a gravitational model
whose degrees of free\-dom solely depend on time. The Einstein equations
for this minisuperspace have been proved equivalent to the
dynamical equations generated by the homogeneous Hamiltonian constraint
of the system when supplemented by a couple of second-class constraints.
We have then imposed all these constraints on the model and elliminated
the unphysical degrees of freedom. The resulting reduced phase space
possesses the symplectic structure of the cotangent bundle over
$I\!\!\!\,R^+\times I\!\!\!\,R^+$. In this reduced phase space,
it is possible to select a complete set of real vari\-ables (observables
in this case) which, under Poisson brackets, form the Lie algebra
$L(T^{\ast}GL(1,I\!\!\!\,R)\times T^{\ast}GL(1,I\!\!\!\,R))$.
We have represented these observables as operators acting on the
vector space of complex functions over $I\!\!\!\,R^+\times I\!\!\!\,R^+$.
The inner product has been determined by promoting the reality conditions
on classical observables to hermiticity requirements on their corresponding
quantum operators. In this way, we have obtained an irreducible
unitary representation of the considered algebra of observables.

We have also revisited the quantization of the diagonal Bianchi type
I (that is, the Kasner model), adopting as close as possible
the language and methods employed in the quantization of the BZ
one-soliton solutions. The quantum theories constructed for the description
of these two models have been shown to be unitarily equivalent,
because the Lie algebras of observables and Hilbert spaces of
physical states associated with these theories
are formally identical. This result can be understood
as an indication supporting the conjecture that, under quantization, the
symmetry between different types of classical spacetimes that underlies in
the BZ transformation with frozen solitonic parameters would translate
into equivalence among the quantum theories which describe
such spacetimes.
\vspace*{.9cm}

{\bf Acknowledgments}

G. A. M. M. was supported by funds provided by DGICYT and the Spanish
Ministry of Education and Science under Contract Adjunct to the Project
No. PB91-0052. N. M. was supported by funds provided by the European
Union HCM Network on Constrained Dynamical Systems under Contract No.
ERBCHRXCT930362.

\end{document}